\documentclass[review]{elsarticle}

\usepackage{lineno}
\biboptions{sort&compress}
\usepackage[colorlinks=true,linkcolor=blue,anchorcolor=blue,citecolor=blue]{hyperref}
\modulolinenumbers[5]
\usepackage{bm}
\usepackage{amsmath}
\usepackage{threeparttable}

\journal{Journal of Nuclear Materials}

\begin{document}

\begin{frontmatter}

\title{Energetics of point defects in aluminum via orbital-free density functional theory}
\author[mymainaddress]{Ruizhi Qiu}
\author[mymainaddress]{Haiyan Lu}
\author[mymainaddress]{Bingyun Ao}
\author[mymainaddress]{Li Huang}
\cortext[mycorrespondingauthor]{Corresponding author}
\ead{lihuang.dmft@gmail.com}
\author[mymainaddress]{Tao Tang}
\author[mymainaddress]{Piheng Chen}
\address[mymainaddress]{Science and Technology on Surface Physics and Chemistry Laboratory,
Mianyang 621908, Sichuan, P.R. China}

\begin{abstract}
The formation and migration energies for various point defects, including vacancies and self-interstitials in aluminum are reinvestigated systematically using the supercell approximation in the framework of orbital-free density functional theory. 
In particular, the finite-size effects and the accuracy of various kinetic energy density functionals are examined. 
The calculated results suggest that the errors due to the finite-size effect decrease exponentially upon enlarging the supercell. 
It is noteworthy that the formation energies of self-interstitials converge much slower than that of vacancy.
With carefully chosen kinetic energy density functionals, the calculated results agree quite well with the available experimental data and those obtained by Kohn-Sham density functional theory which has exact kinetic term. 
\end{abstract}

\end{frontmatter}


\section{Introduction\label{sec:intro}}

Neutron and other energetic particles produced by nuclear reactions usually induce significant changes in the physical properties of irradiated materials. 
Since the radiation defects are often very small and hence not readily accessible to an experimental observation, many atomic-, mesoscopic-, and continuum-level models have been developed to understand the irradiation effect on materials in the past~\cite{Gibson1960,Bai2010,Fu2005,Robinson1974}.
Though great achievements have been obtained with these empirical models, they all make assumptions about the physical laws governing the behaviors of the materials~\cite{Carter2008}.
By contrast, first-principles modeling, which is based on the laws of quantum mechanics, only requires input of the atomic numbers of the elements. 

One of the widely used methods for first-principles modeling is the Kohn-Sham density function theory (dubbed as KS-DFT)~\cite{Hohenberg1964,Kohn1965}.
It has been proven that the KS-DFT method can provide reliable information about the structure of nanoscale defects produced by irradiation, and the nature of short-range interaction between radiation defects, defect clusters, and their migration pathways~\cite{Dudarev2013}.
However, the traditional KS-DFT method is not linear scaling, and at most only a few thousands of atoms can be treated with it using modern supercomputer.
Obviously, it is far away from the requirement of simulating the large atomic system for radiation effect.
The orbital-free density functional theory (dubbed as OF-DFT) method provides another choice for simulating the radiation effect.
Unlike KS-DFT, which uses single-electron orbitals to evaluate the non-interacting kinetic energy, OF-DFT relies on the electron density as the sole variable in the spirit of the Hohenberg-Kohn theorem~\cite{Hohenberg1964} and is significantly less computationally expensive.
The accuracy of OF-DFT depends upon the quality of the used kinetic energy density functional (KEDF), which is usually based on the linear response of a uniform electron gas.
Note that similar to the exchange-correlation density functional (XCDF), the exact form of KEDF is not known except in certain limits.
Currently, the most popular KEDFs are the Wang-Govind-Carter (WGC)~\cite{Wang1999,Wang2001} and Wang-Teter (WT)~\cite{Wang1992} functionals.
Both were designed to reproduce the Lindhard linear response of a free-electron gas~\cite{Lindhard1953}.

In order to apply the OF-DFT method to study radiation defects in realistic materials, it is essential to evaluate the accuracy of various KEDFs. 
The formation and migration energetics of typical point defects in simple metal aluminum are very useful test beds. 
Actually when a new KEDF or formulation was proposed, the vacancy formation energy in Al was always calculated to make a comparison with the experimental value and the KS-DFT results~\cite{Wang1992,Perrot1994,Smargiassi1994,Foley1996,Wang1998,Wang1999,Gavini2007b}.
For example, Foley and Madden~\cite{Foley1996} generalized the WT-KEDF and calculated the relaxed vacancy formation energy in Al on 32- and 108-site cells, and later Jesson \emph{et al.}~\cite{Jesson1997} use the same KEDF to calculate the formation and migration energies of various self-interstitials using 108- and 256-site cell.
But the estimated values differ strongly from the experimental values and the KS-DFT results using the same supercell.
Later Carter's group~\cite{Wang1999} proposed the density-dependent WGC-KEDF and also calculated the vacancy formation energy in Al using a 4-site and 32-site cell.
The estimated value (0.610-0.628) is in very good agreement with the experimental value ($\sim$0.67) and the KS-DFT result in this work (0.626).
Lately the same group~\cite{Ho2007} used various supercell with number of sites up to 1372 to calculate the vacancy formation and migration energies.
But it was found that the calculated results from OF-DFT systematically underestimated the measured values or the KS-DFT results by $\sim$0.2 eV.
Recently Gavini's group~\cite{Gavini2007,Gavini2007b,Radhakrishnan2010,Das2015} proposed the non-periodic real-space formulation for OF-DFT and used this formulation to calculate the vacancy formation energy and also the unrelaxed case. The size effect is also considered here.
The results are in good agreement with those obtained from the OF-DFT calculation using plan-wave basis~\cite{Wang1999,Ho2007}.
Despite those above abundant researches on the properties of point defect in Al, relatively little is known regarding the OF-DFT study of the self-interstitials and the corresponding size effect.
Thus it is necessary to systematically calculate the formation and migration energies of vacancy and various self-interstitials by employing the OF-DFT method with various KEDFs and XCDFs.

The effect of supercell size on the defect energetics is also a major concern of this work.
Due to the periodical boundary condition (PBC) in the routine DFT calculation using the popular plane-wave basis, there may be cross-boundary effects and defect-defect interactions, and therefore a different system other than the intent was studied.
For KS-DFT, the study on the size effect is limited by the small size of the system studied and the imperfection of the Brillouin zone sampling~\cite{Hine2009,Dabrowski2015}, while these disadvantages disappear for OF-DFT.
For example, Ho \emph{et al.}~\cite{Ho2007} found that the formation energy is converged within 3 meV by $4\times4\times4$ supercell and the migration energy is converged within 1 meV at $3\times3\time3$ supercell.
This finite-size errors from the PBC could also been circumvented by using a non-periodic formulation of DFT~\cite{Gavini2007,Gavini2007b}.
The corresponding non-periodic cell-size effect on the energetics of vacancy and divacancy in aluminum using OF-DFT was investigated in References~\cite{Gavini2007b,Radhakrishnan2010,Das2015,Radhakrishnan2016}.
It was revealed that more than 2000 sites are required to obtain a converged value for the divacancy.

In the present work, we employ the OF-DFT method with WGC-KEDF and WT-KEDF to calculate the formation and migration energies of typical point defects in face-centered cubic (fcc) aluminum. 
The simulation cell ranges from $3\times3\times3$ to $14\times14\times14$ supercells.
By comparison, these energies are also calculated by using the KS-DFT method and a $4\times4\times4$ supercell.
The rest of this paper is organized as follows.
The computation methods and details are described in Sec.~\ref{sec:method}. 
In this section, the typical KEDFs used in the calculations are introduced. 
The detailed results are presented and discussed in Sec.~\ref{sec:result}.
Section~\ref{sec:conclusion} serves as a conclusion.

\section{Computational Methods and Details\label{sec:method}}

According to the Hohenberg-Kohn theorem~\cite{Hohenberg1964},
the ground state density of interacting electrons $n({\bm r})$ 
in some external potential $v_{\rm ext}({\bm r})$ determines 
this potential uniquely, 
and the ground state energy $E$ could be obtained variationally,
\begin{eqnarray}
E={\rm min}_{{\tilde n}({\bm r})}E_{\rm tot}[{\tilde n}({\bm r})]
\label{eq:variational}
\end{eqnarray}
under the constraint that the electron density ${\tilde n}({\bm r})$
is non-negative and normalized to the number of electrons $N$.
The minimum is the ground state electron density $n({\bm r})$ for a non-degenerate ground state, which determines all the properties of an electronic ground state.
In the framework of density functional theory, 
the total energy functional in Eq.~(\ref{eq:variational}) could be expressed as
\begin{eqnarray}
E_{\rm tot}[n({\bm r})]
= T_s[n({\bm r})]+E_{\rm xc}[n({\bm r})]
+E_{\rm H}[n({\bm r})]
+\int v_{\rm ext}({\bm r}) n({\bm r})d{\bm r}
\label{eq:energy}
\end{eqnarray}
where $T_s[n({\bm r})]$, $E_{\rm H}[n({\bm r})]$ and $E_{\rm xc}[n({\bm r})]$ represent
the kinetic energy of the ground state
of the non-interacting electrons with density $n({\bm r})$,
the Hartree electrostatic energy
and the exchange-correlation energy, respectively.

In the framework of OF-DFT method, the kinetic term $T[n]$ is approximated using KEDF.
Here we adopt the most accurate functionals available, i.e.,
the Wang-Govind-Carter (WGC)~\cite{Wang1999,Wang2001}
and Wang-Teter (WT)~\cite{Wang1992} KEDFs.
They consist of the Thomas-Fermi
(TF) functional~\cite{Thomas1927,Fermi1927}, the von Weizs\"acker (vW) functional
~\cite{Weizsaecker1935}, and a linear response term.
In reduced units, 
the TF-KEDF is given by
\begin{eqnarray}
\label{eq:TF}
T_{\rm TF}[n({\bm r})]
=\int n({\bm r})\frac{3}{10}k_{\rm F}^2[n({\bm r})]d{\bm r},
\end{eqnarray}
where $k_{\rm F}[n]=(3\pi^2 n)^{1/3}$ is the Fermi wave vector
of a uniform electron gas of density $n$ and $\textstyle{\frac{3}{10}}k_{\rm F}^2[n]$
is the mean kinetic energy per electron of such a gas.
The vW-KEDF reads
\begin{eqnarray}
\label{eq:vW}
T_{\rm vW}[n({\bm r})]
=\frac{1}{8}\int \frac{\left\vert\nabla n({\bm r})\right\vert^2}{n({\bm r})} d{\bm r},
\end{eqnarray}
which could be obtained from a single-orbital occupied system.
Response functions of an electronic system is of vital importance
to understand its physical properties, 
and for a non-interacting electron gas,
the correct linear response behavior was derived analytically by
Lindhard~\cite{Lindhard1953}.
In Lindhard's theory,
the static electric susceptibility in reciprocal space is given by
\begin{eqnarray}
\chi_{\rm Lind}({\bm q}) = 
-\frac{k_{\rm F}[n_0]}{\pi^2}
\left(\frac{1}{2}+\frac{1-\eta^2}{4\eta}
\ln\left\vert\frac{1+\eta}{1-\eta}\right\vert\right),
\quad \eta = \left\vert\frac{\bm q}{2 k_{\rm F}}\right\vert
\end{eqnarray}
where $n_0$ is the average electron density.
But for the TF-KEDF and vW-KEDF, the corresponding
susceptibility functions are given by $\chi_{\rm TF}=-k_{\rm F}[n_0]/\pi^2$
and $\chi_{\rm vW}=\chi_{\rm TF}/(3\eta^2)$, respectively.
In order to remedy this, 
one must introduce a linear response term and then 
the resulting KEDF could be given by
\begin{eqnarray}
\label{eq:wgc}
T^{\alpha,\beta}_{\rm WT}n
=T_{\rm TF}[n]+T_{\rm vW}[n]+\iint 
\left\{n({\bm r})\right\}^{\alpha}
w({\bm r},{\bm r}')
\left\{n({\bm r}')\right\}^{\beta} d{\bm r}d{\bm r}'.
\end{eqnarray}
For the WT-KEDF~\cite{Wang1992},
the linear response kernel 
$w({\bm r},{\bm r}')$
takes the local form, 
\begin{eqnarray}
w({\bm r},{\bm r}')=w_{\alpha,\beta}({\bm r}-{\bm r}')
={\cal F}^{-1}\left[
-\frac{\chi_{\rm Lind}^{-1}({\bm q})-\chi_{\rm vW}^{-1}({\bm q})-\chi_{\rm TF}^{-1}}
{2\alpha\beta n_0^{\alpha+\beta-2}}\right]({\bm r}-{\bm r}'),
\end{eqnarray}
where ${\cal F}[\cdot]({\bm q})$ denotes the Fourier transform.
While for the WGC-KEDF~\cite{Wang1999,Wang2001},
the kernel takes the non-local form, 
\begin{eqnarray}
w({\bm r},{\bm r}')=
w_{\alpha,\beta,\gamma}\left(\left[\frac{k_{\rm F}^\gamma[n({\bm r})]
+k_{\rm F}^\gamma[n({\bm r}')]}{2}\right]^{1/\gamma}
\left\vert{\bm r}-{\bm r}'\right\vert\right),
\end{eqnarray}
and the exact functional form
could be determined by solving a second-order differential
equation~\cite{Wang1999,Wang2001}.
Actually, the WGC-KEDF kernel is evaluated by performing a Taylor series
of $n({\bm r})-n^\star$ with $n^\star$ being a reference density and usually
chosen to be the average density $n_0$.
In this work, the Taylor series expansion are evaluated up to second order.
The exponent $\alpha$ and $\beta$ could be 
treated as fitting parameters~\cite{Wang1992,Pearson1993,Foley1994,Smargiassi1994,Smargiassi1995,Foley1996,Jesson1997}
or determined from an asymptotic analysis
~\cite{Wang1998,Wang1999b,Wang2001b}.
In the following, we choose $\alpha=\textstyle{\frac{5}{6}}+\textstyle{\frac{\sqrt{5}}{6}}$ and
$\beta=\textstyle{\frac{5}{6}}-\textstyle{\frac{\sqrt{5}}{6}}$ for WGC-KEDF and $\alpha=\beta=\textstyle{\frac{5}{6}}$ for WT-KEDF.
The exponent $\gamma$ is a material-specific adjustable parameter.
According to the literatures, $\gamma=2.7$ is found to be optimal for Al~\cite{Wang1999,Wang2001}.

In the KS-DFT method~\cite{Kohn1965}, 
the KEDF could be expressed in terms of the Kohn-Sham orbitals $\psi_k({\bm r})$,
\begin{eqnarray}
\label{eq:kin}
T_s[n({\bm r})]
=-\frac{1}{2}\sum_{k=1}^N f_k
\int \psi^*_k({\bm r}) \nabla^2\psi_k({\bm r}) d{\bm r},
\end{eqnarray}
where the sum is over the $N$ lowest-energy orbitals $\psi_k$ and $f_k$ represents the corresponding occupancy in the orbital $\psi_k$.
The Kohn-Sham orbitals satisfies 
$n({\bm r})=\sum_{k=1}^N\left\vert\psi_k({\bm r})\right\vert^2$. In principles, the expression in Eq.~(\ref{eq:kin}) is exact, while those in Eq.~(\ref{eq:TF}), (\ref{eq:vW}), and (\ref{eq:wgc}) are not.

For KS-DFT, the commonly used pseudopotential schemes, which represent $v_{\rm ext}({\bm r})$ in Eq.~(\ref{eq:energy}), are usually non-local and could be expressed by using Kohn-Sham orbitals.
However, since there is no orbitals in OF-DFT, the so-called bulk-derived local pseudopotential (BLPS)~\cite{Zhou2004,Huang2008} is used here for describing the potential $v_{\rm ext}({\bm r})$.
Both in KS-DFT and OF-DFT, the exchange-correlation energy functionals $E_{\rm xc}[n]$ in Eq.~(\ref{eq:energy}) are described by using the local density approximation (LDA)~\cite{Perdew1981} and the generalized gradient approximation (GGA) of Perdew, Burke and Ernzerhof (PBE)~\cite{Perdew1996}.

In the present work, we use the PROFESS code to do the OF-DFT calculations~\cite{Ho2008,Hung2010,Chen2015}.
Multiple simulation box sizes, ranging from $3\times3\times3$ to $14\times14\times14$ supercells, were tested here.
Let us denote the supercell size as $N_s$, then the corresponding number of atoms in the perfect supercell of fcc-Al is given by $4N_s^3$ which ranges from 108 to 10976.
For electron density optimization, the kinetic energy cutoff is set to 1200 eV, and  the square root truncated Newton minimization method is used with energy convergence threshold being 2.72$\times$10$^{-5}$ eV.
For ion relaxation, dynamical boundary conditions are employed, allowing cell volume and cell shape to change, along with the relaxation of atom positions.
In addition, both the conjugate gradient and quickmin algorithm are used with convergence threshold for the maximum force component on any atom being 2.57$\times$10$^{-3}$ eV$\cdot$\AA$^{-1}$.

For comparison, we also calculated the point defect energies in a $4\times4\times4$ supercell using the VASP code which implemented the KS-DFT method~\cite{Kresse1996}.
The projector augmented wave (PAW)~\cite{Blochl1994,Kresse1999} pseudopotential for Al with the $3s^22p^1$ valence electronic configuration is chosen.
The kinetic energy cutoff is taken as 400 eV and the $3\times3\times3$ $k$ point meshes based on Monkhorst-Pack scheme~\cite{Monkhorst1976} are adopted.
Methfessel and Paxton’s smearing method~\cite{Methfessel1989} of the first order is used with a width of 0.1 eV to determine the partial occupancies for each Kohn-Sham orbitals.
Relaxations are performed by employing the dynamical boundary conditions and using the conjugate gradient and quasi-Newton algorithm with a convergence criterion of 1 meV with regards to the total free energy of the system. 

\section{Results and Discussion\label{sec:result}}

\subsection{Equilibrium lattice constant}

\begin{table}[th]
\begin{center}
\caption{
Optimized lattice parameters $a_0$ of fcc-Al obtained by the KS-DFT and OF-DFT methods with various XCDFs and KEDFs. The experimental value extrapolated to 0 K is also shown for comparison.
}
\label{tab:lattice}
\small
\begin{tabular}{cccc}
\hline
Method & KEDF & XCDF & $a_0$~(\AA) \\
\hline
KS-DFT & -   & LDA & 3.9830 \\
KS-DFT & -   & GGA & 4.0387 \\
OF-DFT & WGC & LDA & 3.9725 \\
OF-DFT & WGC & GGA & 4.0579 \\
OF-DFT & WT  & LDA & 3.9849 \\
OF-DFT & WT  & GGA & 4.0676 \\
KS-DFT~\cite{Narasimhan1995,Huang2008,Das2015} & -   & LDA & 3.945-3.968 \\
KS-DFT~\cite{Zhuang2016} & -   & GGA & 4.063 \\
OF-DFT~\cite{Carling2003b} & Perrot~\cite{Perrot1994} & LDA & 4.06 \\
OF-DFT~\cite{Carling2003b} & SM~\cite{Smargiassi1994} & LDA & 3.96 \\
OF-DFT~\cite{Carling2003b} & WT~\cite{Wang1992} & LDA & 4.04 \\
OF-DFT~\cite{Carling2003b} & WGC~\cite{Wang1999,Wang2001} & LDA & 4.03-4.04 \\
OF-DFT~\cite{Radhakrishnan2010} & WGC~\cite{Wang1999,Wang2001} & LDA & 4.022 \\
OF-DFT~\cite{Das2015} & WGC & LDA & 3.973 \\
OF-DFT~\cite{Zhuang2016} & WGC & GGA & 4.039 \\
OF-DFT~\cite{Jesson1997} & FM~\cite{Foley1996}  & LDA & 4.0270 \\
Experiment~\cite{Bandyopadhyay1978} & - & - & 4.0315 \\
\hline
\end{tabular}
\end{center}
\end{table}

For the bulk properties and the energies of several competing phases of bulk Al, Shin \emph{et al.}~\cite{Shin2009} has already made detailed comparison between the OF-DFT and KS-DFT methods with various exchange-correlation density functionals (XCDFs) and KEDFs. 
It was found that the OF-DFT results accurately reproduced those by the KS-DFT method.
This manifests that the OF-DFT method with reliable BLPS and KEDF is an accurate simulating tool for perfect crystal.
Before introducing the point defect, here we preformed a structural relaxation on the perfect supercell for fcc-Al to figure out the equilibrium lattice constant $a_0$. 
In Table~\ref{tab:lattice} we compare the lattice constants of fcc-Al obtained from various OF-DFT and KS-DFT calculations with the experimental value and the previous theoretcial results.
The experimental value shown in Table~\ref{tab:lattice} is obtained by extrapolating to 0 K using the polynomial proposed in Ref.~\cite{Bandyopadhyay1978}.
Clearly, all the lattice constants obtained from OF-DFT are accurate enough to perform the further investigation of defect energetics.

%
\subsection{Formation energy of vacancy}

\begin{figure}[t]
\begin{center}
\includegraphics[width=0.8\textwidth]{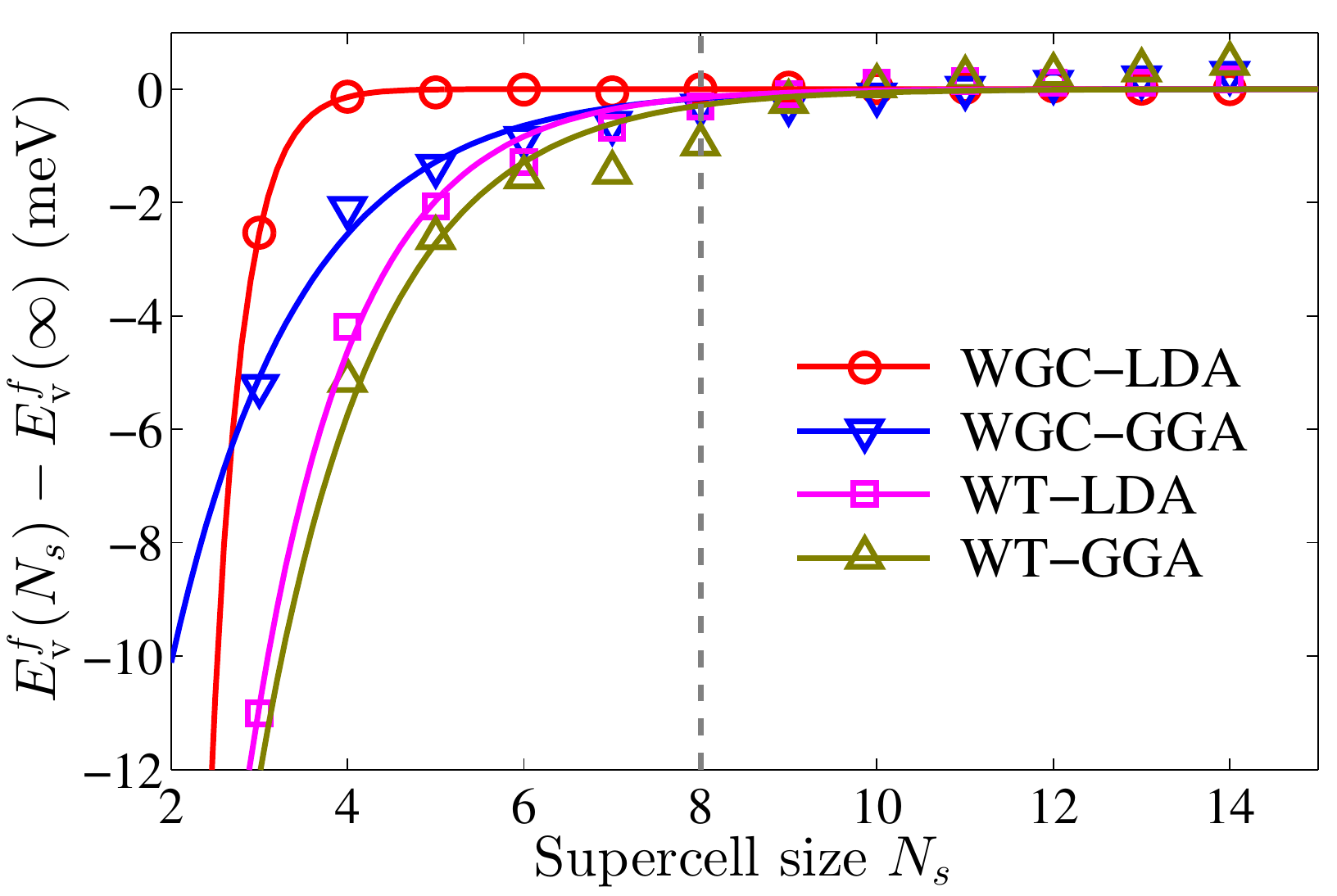}
\caption{
The supercell size dependence of vacancy formation energy under different combinations of XCDFs and KEDFs.
The symbols indicate the calculated values from OF-DFT calculation and the lines represent the fitting curves.
The dashed line indicates the optimal supercell size.
\label{fig:vacancy}
}
\end{center}
\end{figure}

\begin{table}[t]
\begin{center}
\caption{
Vacancy formation energies of fcc-Al $E^{f}_{\rm v}$ obtained by the KS-DFT and OF-DFT methods with different XCDFs and KEDFs. 
The experimental value at ``moderate" temperature and the previous theoretical results are also shown for comparison.
}
\label{tab:vacancy}
\vspace{0.1in}
\begin{tabular}{cccc}
\hline
Method & KEDF & XCDF & $E^f_{\rm v}$~(eV) \\
\hline
KS-DFT & -   & LDA & 0.7291 \\
KS-DFT & -   & GGA & 0.6646 \\
OF-DFT & WGC & LDA & 0.7947 \\
OF-DFT & WGC & GGA & 0.7213 \\
OF-DFT & WT  & LDA & 1.3456 \\
OF-DFT & WT  & GGA & 1.3001 \\
KS-DFT~\cite{Chetty1995,Turner1997,Wang1999,Carling2000,Baskes2001,Carling2003,Uesugi2003,Lu2005,Ho2007,Iyer2014} 
& -   & LDA & 0.66-0.73 \\
KS-DFT~\cite{Carling2000,Carling2003,Lu2005} & -   & GGA & 0.54-0.55 \\
OF-DFT~\cite{Foley1996} & FM~\cite{Foley1996} & LDA & 0.29 \\
OF-DFT~\cite{Wang1999,Ho2007,Gavini2007b,Radhakrishnan2010} & WGC & LDA & 0.48-0.72 \\
OF-DFT~\cite{Ho2007} & WGC & GGA & 0.387\\
Experiment~\cite{Ullmaier1991} & - & - & 0.67$\pm$0.03 \\
\hline
\end{tabular}
\end{center}
\end{table}

Typical defects were then introduced into the fully relaxed supercells, and the structural relaxation without any symmetry constraints was performed again for the given supercell to calculate formation energies.
For a single vacancy defect, it is created by eliminating one central atom in the supercell and the corresponding formation energy is defined as~\cite{Jesson1997}
\begin{eqnarray}
E_{\rm v}^f = E_{(n-1){\rm Al}}
-\frac{n-1}{n} E_{n{\rm Al}},
\end{eqnarray}
where $E_{(n-1){\rm Al}}$ is the total energy for the supercell containing an $(n-1)$Al atoms and one vacancy, and $E_{n{\rm Al}}$ is the total energy of a perfect aluminum supercell containing $n$ Al atoms.
For an Al interstitial defect, it is generated by adding a single Al atom into the supercell in different interstitial positions around the center. The corresponding formation energy is given by
\begin{eqnarray}
E_{\rm i}^f = E_{(n+1){\rm Al}}
-\frac{n+1}{n} E_{n{\rm Al}},
\end{eqnarray}
where $E_{(n+1){\rm Al}}$ is the total energy for the supercell containing $(n)$ Al atoms at lattice sites and one interstitial aluminum atom.
Since the full relaxation is performed on the prefect and defect-containing system, $E_{\rm v}^f$ and $E_{\rm i}^f$ are the formation energy at constant zero pressure~\cite{Gillan1989,Ho2007,Iyer2014}.

We at first try to test the size dependence of vacancy formation energy $E_{\rm v}^f$.
The results are shown in Figure~\ref{fig:vacancy}.
The $x$-axis is the size of supercell $N_s$.
For a specified defect, the distance between it and its image defect located in the neighboring cell is given by $d=N_s a_0$.
Surely, the increment of $N_s$ leads to larger $d$, and thus reduces the defect-defect interaction. As a consequence, the formation energy approaches the converged value asymptotically upon increasing $N_s$.
This tendency is clearly seen in Figure~\ref{fig:vacancy} for different computational methods.
In particular, the WGC-KEDF converges more rapidly than the WT-KEDF, and the LDA-XCDF performed better than the GGA-XCDF.
Note that the WGC-KEDF plus LDA-XCDF performed extremely well and was chosen in the previous literatures~\cite{Ho2007,Radhakrishnan2010,Das2015}.

In order to model the convergent tendency, we used a simple exponential function 
\begin{eqnarray}
E_{\rm v}^f = A\;\exp\left(-B\; N_s\right) + C
\label{eq:fit}
\end{eqnarray}
to fit the vacancy formation energy.
The fitting curves are also shown in Figure~\ref{fig:vacancy} and apparently, the fit is very well.
Since the fit always give a positive value of $B$ in equation~(\ref{eq:fit}), the value of $C$ may be regarded as the formation energy in the thermodynamic limit ($N_s \rightarrow \infty$).
Here we realized that finite-size errors decrease exponentially as $N_s$ is increased, and finite-size scaling could give a reliable value.

The convergence of $E_{\rm v}^f$ with respect to $N_s$ was already found in the previous literatures~\cite{Ho2007,Radhakrishnan2010,Iyer2014,Das2015}.
For WGC-KEDF, Ho \emph{et al.}~\cite{Ho2007} found that the formation energy is converged within 3 meV by $4\times4\times4$ supercell, which is also confirmed by Gavini's group~\cite{Radhakrishnan2010,Iyer2014,Das2015} and our calculation (see Figure~\ref{fig:vacancy}).
But for WT-KEDF, the convergence of $E_{\rm v}^f$ with respect to $N_s$ is not as quick as that for WGC-KEDF.
For a convergence criterion of 1 meV, a supercell size of $8\times8\times8$ is required for the vacancy formation energy.

The vacancy formation energies of fcc-Al obtained with the OF-DFT method using various XCDFs and KEDFs are collected and summarized in Table~\ref{tab:vacancy}. 
For comparison, the related KS-DFT values for a $4\times4\times4$ supercell, the previous theoretical results from KS-DFT and OF-DFT, and the experimental value are also shown.
Clearly, the calculated formation energies from the OF-DFT method with WGC-KEDF are close to that of KS-DFT.
However, since the kernel in WT-KEDF is density-independent, it isn't suitable for systems where the electron density varies rapidly, such as vacancies or surfaces~\cite{Wang1999,Wang2001}. 
It is not surprised that the results obtained with the WT-KEDF deviate from the others apparently.
In addition, both the values from KS-DFT and OF-DFT with WGC-KEDF are in good agreement with the experimental value.

\begin{figure}[t]
\begin{center}
\includegraphics[width=0.9\textwidth]{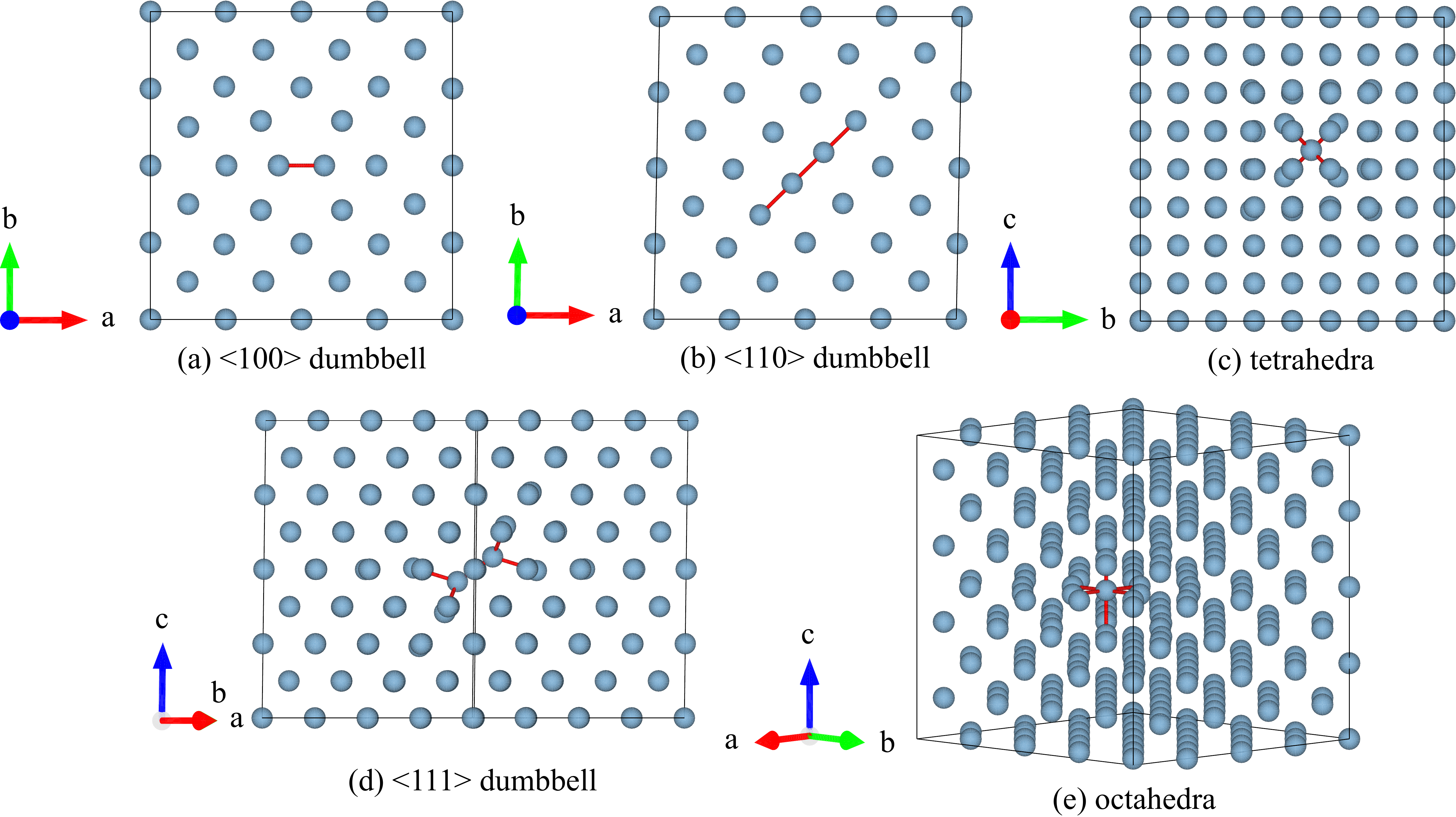}
\caption{The geometry configurations of self-interstitials in fcc-Al within a $4\times4\times4$ supercell. (a) $\langle 100\rangle$ dumbbell; (b) $\langle 110\rangle$ dumbbell; (c) tetrahedron; (d) $\langle111\rangle$ dumbbell; (e) octahedron. Note that slices are displayed in (a) and (b).}
\label{fig:interstitial}
\end{center}
\end{figure}

\subsection{Formation energies of self-interstitials}

\begin{figure}[t]
\begin{center}
\includegraphics[width=0.8\textwidth]{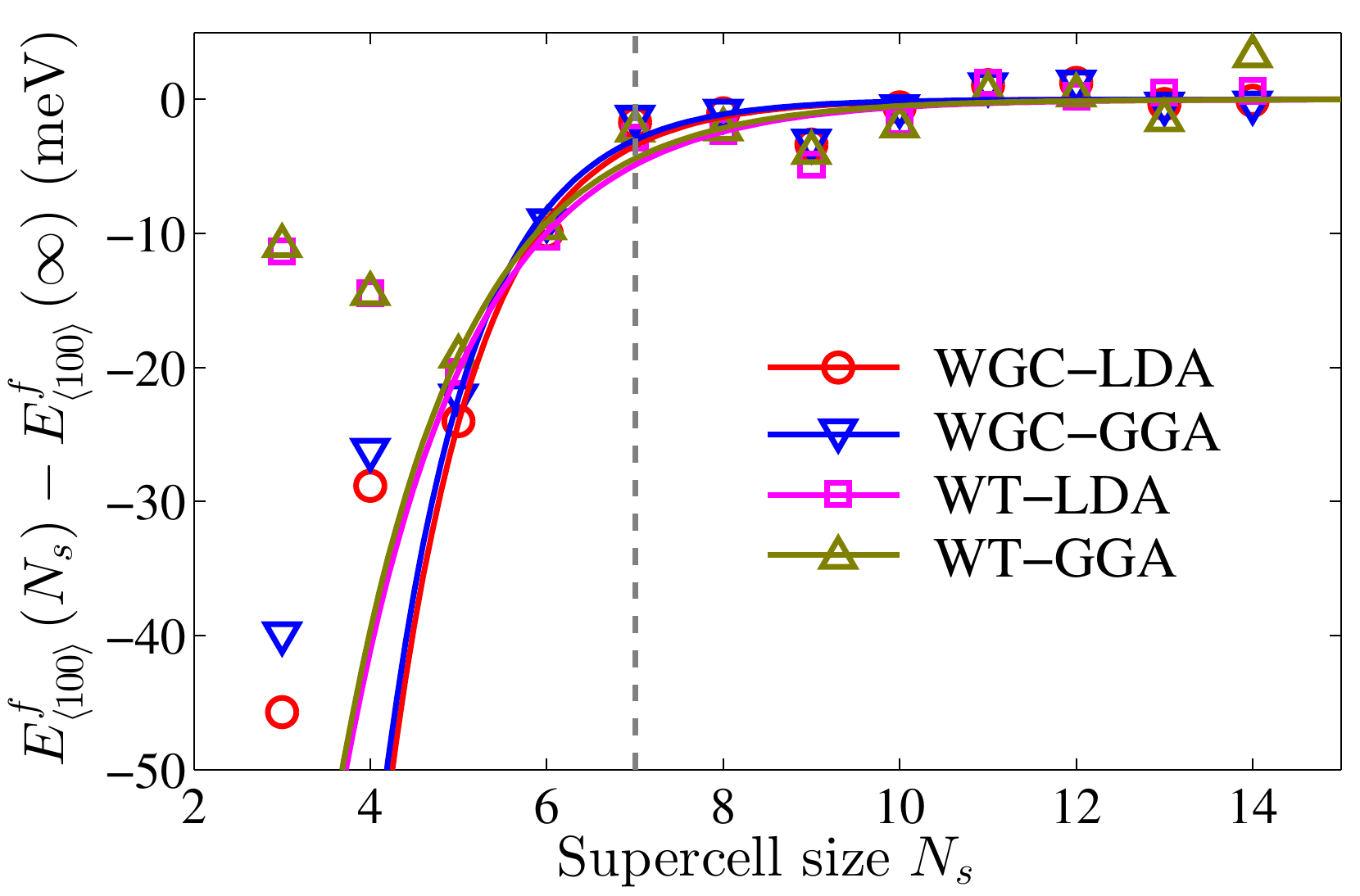}
\caption{
The supercell size dependence of $\langle100\rangle$ dumbbell interstitial formation energy under different combinations of XCDFs and KEDFs.
The symbols indicate the calculated values from OF-DFT calculation and the lines represent the fitting curves.
The dashed line indicates the optimal supercell size.
\label{fig:sia100}
}
\end{center}
\end{figure}

Now let's turn to the self-interstitials in fcc-Al. 
Five types of self-interstitials, including $\langle 100\rangle$ dumbbell, $\langle 110\rangle$ dumbbell, $\langle111\rangle$ dumbbell, octahedral and tetrahedral interstitials, are considered in the present work.
The corresponding formation energies are denoted as $E^f_{\langle100\rangle}$, $E^f_{\langle110\rangle}$, $E^f_{\langle111\rangle}$, $E^f_{\rm oct}$, and $E^f_{\rm tet}$, respectively.
The relaxed geometry configurations from KS-DFT are shown in Figure~\ref{fig:interstitial}. 
Those configurations obtained by the OF-DFT method are very similar and wouldn't be shown here.

Being analogous to the vacancy formation energies as discussed before, the interstitial formation energies also show a convergent behavior with respect to the supercell size.
As an example, the formation energies of $\langle 100\rangle$ dumbbell interstitial $E^f_{\langle100\rangle}$ 
as a function of $N_s$ and the fitting curves are shown in Figure~\ref{fig:sia100}.
Note that the values obtained from small supercell sizes deviate from the fitting curves and were not used in the fitting process.
The convergence speed of $E^f_{\langle100\rangle}$ is much slower than that of $E^f_{\rm v}$.
For a rough convergence criterion of 5 meV, a supercell size of $7\times7\times7$ is required for the formation energy of the $\langle100\rangle$ dumbbell interstitial.
If we only focus on for the convergent behavior, the WGC-KEDF does not perform better than the WT-KEDF, which differs from the case of vacancy formation energy.
Here the $C$ values in Eq.~(\ref{eq:fit}) from the exponential fit are also taken as the formation energies in the thermodynamic limit.

Table~\ref{tab:interstitial} shows the interstitial formation energies of fcc-Al obtained from the OF-DFT and KS-DFT methods with various density functionals.
The pervious theoretical results from KS-DFT~\cite{Iyer2014} and OF-DFT~\cite{Jesson1997} using $3\times3\times3$ supercell, and the experimental value were also shown for comparison.
Note that the experimental value was estimated from the experimental value of Frenkel pair formation energy, 3.7 eV~\cite{Ullmaier1991} and the vacancy formation energy, 0.67$\pm$0.03 eV~\cite{Ullmaier1991}.
We met convergence problems when we performed OF-DFT calculations with WGC-KEDF except for the $\langle 100\rangle$ dumbbell configuration.
In order to cure this trouble, we repeated the OF-DFT calculations by setting $\partial^2 w/\partial n^2 = 0$ in the WGC-KEDF (denoted as WGC* in the following).
The calculated results show that the modified WGC-KEDF leads to slightly larger formation energies.

\begin{threeparttable}[th]
\begin{center}
\caption{
Interstitial formation energies $E^{f}_{\rm I}$ (in units of eV) obtained from the KS-DFT and OF-DFT method with various XCDFs and KEDFs. 
The subscripts $\langle 100\rangle$, $\langle 110\rangle$, $\langle 111\rangle$, ${\rm oct}$ and ${\rm tet}$ mean the corresponding geometry configurations as shown in Figure~\ref{fig:interstitial}.
The experimental value was estimated from the formation energies of Frenkel pair and vacancy, and therefore should be treated as an average.
}
\label{tab:interstitial}
\begin{tabular}{llllllll}
\hline
Method & KEDF & XCDF & $E^f_{\langle 100\rangle}$ & $E^f_{\langle 110\rangle}$ & $E^f_{\langle 111\rangle}$ & $E^f_{\rm oct}$ & $E^f_{\rm tet}$ \\
\hline
KS-DFT & -   & LDA & 2.6073 & 2.9809 & 3.1821 & 2.9485 & 3.2941 \\
KS-DFT & -   & GGA & 2.4597 & 2.7508 & 3.0183 & 2.7945 & 3.1072 \\
OF-DFT & WGC & LDA          & 2.5862 &  &  &  &  \\
OF-DFT & WGC\tnote{*} & LDA  & 2.6031 & 2.8854 & 3.0986 & 2.8728 & 3.1466  \\
OF-DFT & WGC & GGA          & 2.3925 &  &  &  &  \\
OF-DFT & WGC\tnote{*} & GGA  & 2.4200 & 2.6856 & 2.8784 & 2.6724 & 2.9217 \\
OF-DFT & WT  & LDA          & 2.4257 & 2.7000 & 2.8507 & 2.6672 & 2.8749 \\
OF-DFT & WT  & GGA          & 2.2943 & 2.5422 & 2.6789 & 2.5302 & 2.6998 \\
KS-DFT~\cite{Iyer2014} & -   & LDA & 2.9  & - & - & 3.1 & 3.5 \\
OF-DFT~\cite{Jesson1997} & FM~\cite{Foley1996} & LDA & 1.579 & 1.869 & 1.959 & 1.790 & 1.978 \\
Exp.~\cite{Ullmaier1991} & - & - &  &  & 3.0$^\dag$  &  & \\
\hline
\end{tabular}
\begin{tablenotes}
\item{*} Here $\partial^2 w/\partial n^2$ in WGC-KEDF is set to be zero.
\item{$^\dag$} The average value of formation energy, not the value of $E^f_{\langle 111\rangle}$.
\end{tablenotes}
\end{center}
\end{threeparttable}

It is surprising to found that the OF-DFT results with WGC-KEDF reproduce extremely well the KS-DFT values with errors being less than 0.15 eV.
In addition, we find that for LDA-XCDF, both OF-DFT and KS-DFT calculations yield, 
\begin{eqnarray}
E^f_{\langle 100\rangle} < E^f_{\rm oct} < E^f_{\langle 110\rangle} <  E^f_{\langle 111\rangle}  <  E^f_{\rm tet}.
\nonumber
\end{eqnarray}
It is consistent with the experimental finding that the $\langle 100\rangle$ dumbbell is the favorite interstitial configuration~\cite{Ehrhart1973,Ehrhart1974,Robrock1976}.
In addition, this order was the same as that of OF-DFT calculation with the FM-KEDF~\cite{Jesson1997} and that of KS-DFT calculation~\cite{Iyer2014}.
For GGA-XCDF, the only exception is $E_{\langle110\rangle} < E_{\rm oct}$ for the KS-DFT case.

The results from KS-DFT and OF-DFT with WGC-KEDF are both consistent with the experimental value 3.0 eV~\cite{Ullmaier1991}.
But the results from OF-DFT plus WT-KEDF and FM-KEDF deviate from the experimental value since all the calculated values for different geometries are smaller than 3.0 eV.
The results~\cite{Jesson1997} from FM-KEDF differ strongly from our values, and also the KS-DFT and experimental data. 
Those discrepancies may stem from the local pseudopotentials.
For KS-DFT, the estimated values in our work is in agreement with the previous theoretical results and the difference may result from the different supercells and computational parameters.

\subsection{Migration energies of vacancy and $\langle100\rangle$ dumbbell}

\begin{figure}[t]
\begin{center}
\includegraphics[width=0.8\textwidth]{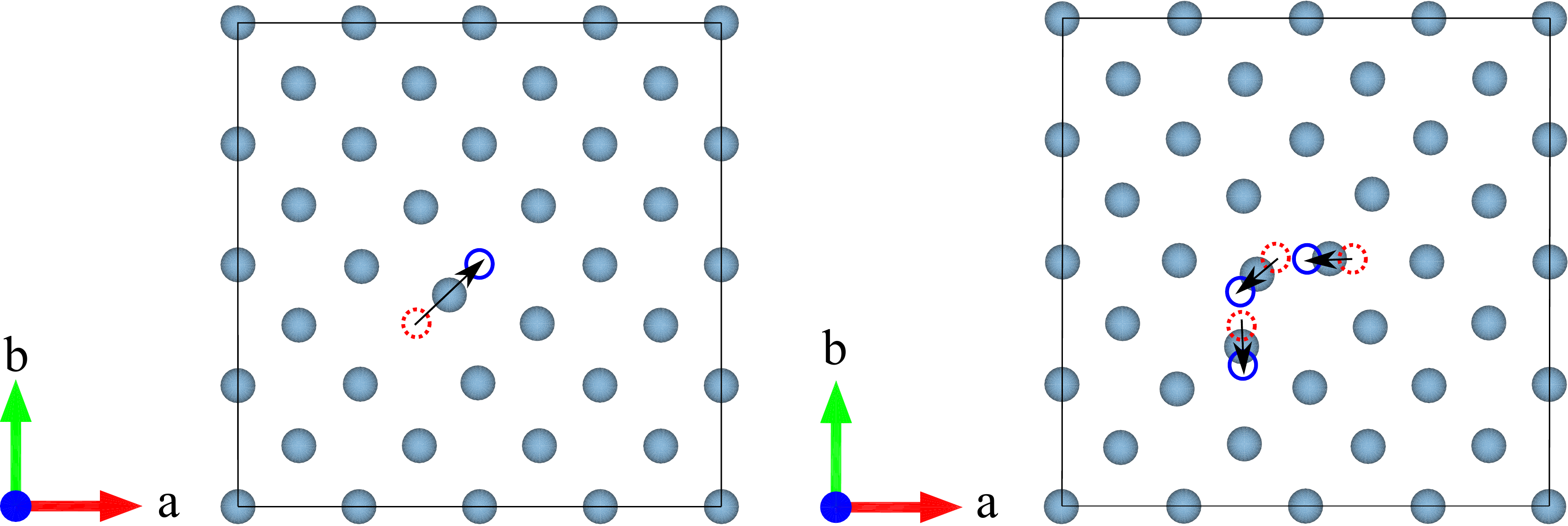}
\caption{
The intermediate atomic configurations for (a) vacancy migration and (b) $\langle100\rangle$ dumbbell migration.
For the migration atoms, the arrows, the red dotted circles, and the blue circles schematically represent the migration direction, and the positions before and after the migration, respectively.
}
\label{fig:migration}
\end{center}
\end{figure}

\begin{figure}[t]
\begin{center}
\includegraphics[width=0.8\textwidth]{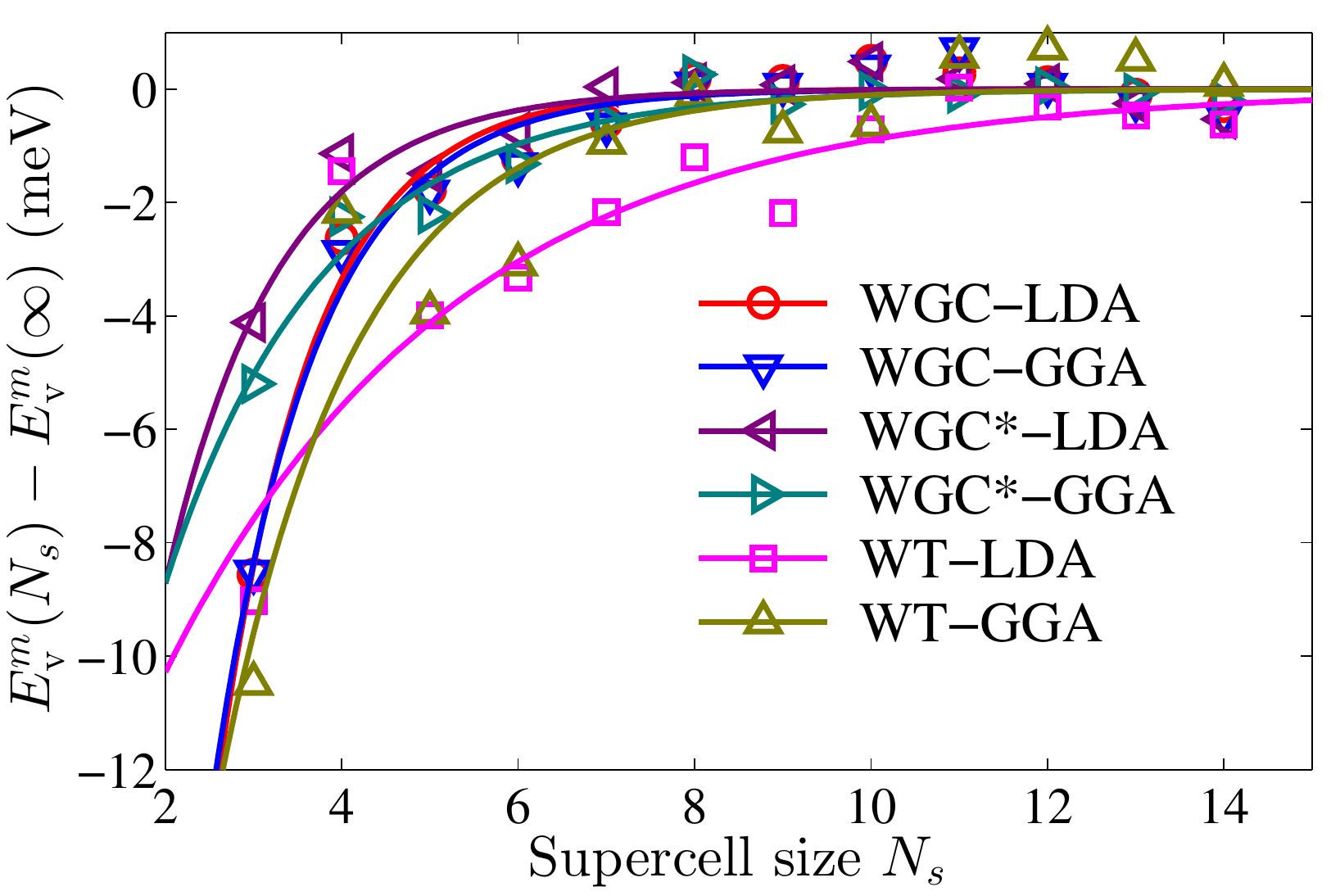}
\caption{
The supercell size dependence of vacancy migration energy under different combinations of XCDFs and KEDFs.
The symbols indicate the calculated values from OF-DFT calculations, while the lines represent the fitting curves.}
\label{fig:emigration}
\end{center}
\end{figure}

The general migration pathway of a vacancy is that one atom moves towards the adjacent vacancy, eliminating the vacancy and forming a new vacancy.
For the $\langle 100\rangle$ dumbbell, the easiest migration pathway was already discussed by Jesson \emph{et al.}~\cite{Jesson1997}.
In this case, one dumbbell atom moves towards an adjacent interstitial site and the other dumbbell atom returns to the lattice site, forming a new $\langle010\rangle$ or $\langle 001\rangle$ dumbbell.
The intermediate configurations for migration of vacancy and $\langle 100\rangle$ dumbbell are depicted in Figure~\ref{fig:migration}.
In order to calculate the migration energy, i.e., the potential barrier in the migration pathway, all of the possible intermediate configurations are relaxed at first, and then their energies are calculated. 
Then the migration energy is given by the difference between the maximum total energy of intermediate configuration and the total energy of initial configuration.
For KS-DFT with LDA-XCDF, the intermediate atomic configurations for migration of vacancy and $\langle 100\rangle$ dumbbell with maximum total energy are shown in Figure~\ref{fig:migration}.

The migration energies also show a convergent behavior with respect to the supercell size $N_s$.
In Figure~\ref{fig:emigration}, the supercell size dependence of vacancy migration energy $E^m_{\rm v}$ are shown.
The convergence speed of $E^m_{\rm v}$ is not as good as that of $E^f_{\rm v}$.
Especially, it is difficult to obtain converged $E^m_{\rm v}$ with the OF-DFT method using WT-KEDF, which confirms again that the WT-KEDF isn't suitable for describing systems with vacancy defects.
In addition, the convergence of $E^m_{\langle100\rangle}$ is not as good as that of $E^m_{\rm v}$.
Thus in the following OF-DFT calculations, we just use the results from the $14\times14\times14$ supercell.

Table~\ref{tab:migration} shows the migration energies of vacancy and $\langle100\rangle$ dumbbell obtained from the OF-DFT and KS-DFT methods.
We also show the experimental values and the available theoretical results for comparison.
The vacancy migration energy $E^m_{\rm v}$ obtained by the OF-DFT method with WGC-KEDF reproduced well that from the KS-DFT method and also the experimental value.
As illustrated above, due to the density-independent kernel, the $E^m_{\rm v}$ values from OF-DFT with WT-KEDF can't reproduce that from KS-DFT and also in bad agreement with the experimental values.

For the $\langle 100\rangle$ dumbbell migration energy $E^m_{\langle100\rangle}$, the OF-DFT results differ from the results from KS-DFT.
However, the $E^m_{\langle100\rangle}$ values from the OF-DFT method are in close agreement with the experimental results while the KS-DFT results are not.
In addition, the deviations of the OF-DFT results from the previous theoretical results are acceptable.

\begin{center}
\begin{threeparttable}[th]
\caption{
The migration energies (in units of eV) of vacancy and $\langle100\rangle$ dumbbell obtained from the OF-DFT and KS-DFT methods with various density functionals.
}
\label{tab:migration}
\begin{tabular}{lllll}
\hline
Method & KEDF & XCDF & $E^m_{\rm v}$ & $E^m_{\langle 100\rangle}$ \\
\hline
KS-DFT & -   & LDA & 0.5234 & 0.2130 \\
KS-DFT & -   & GGA & 0.4904 & 0.2179  \\
OF-DFT & WGC & LDA                  & 0.5846 & -\tnote{$^\ddagger$}  \\
OF-DFT & WGC\tnote{*} & LDA  & 0.6251 & 0.1168  \\
OF-DFT & WGC & GGA                  & 0.5569 & -\tnote{$^\ddagger$}  \\
OF-DFT & WGC\tnote{*} & GGA  & 0.5964 & 0.1101 \\
OF-DFT & WT  & LDA                  & 0.3135 & 0.1119 \\
OF-DFT & WT  & GGA                  & 0.3041 & 0.1041 \\
OF-DFT~\cite{Jesson1997} & FM~\cite{Foley1996} & LDA & - & 0.084 \\
OF-DFT~\cite{Ho2007} & WGC & LDA & 0.42 & - \\
Exp.~\cite{Ullmaier1991} & - & - & 0.61$\pm$0.03 & 0.115$\pm$0.025 \\
\hline
\end{tabular}
\begin{tablenotes}
\item{*} Here a modified WGC-KEDF is used with $\partial^2 w/\partial n^2 = 0$.
\item{$^\ddagger$} In this scheme, the electron density does not converge.
\end{tablenotes}
\end{threeparttable}
\end{center}

\section{Conclusion}
\label{sec:conclusion}

In summary, the OF-DFT method combined with the WGC-KEDF and WT-KEDF was employed to calculate the formation energies and migration energies of typical point defects in fcc-Al supercell up to 10976 atoms.
The finite-size errors arising from the supercell approximation are examined and could be corrected for using finite-size scaling methods.
The convergence of interstitials formation energies is much slower than that of vacancy formation energy.
And our cell-size study of $\langle100\rangle$ dumbbell interstitial has shown that it is converged within 5 meV by a $7\times7\times7$ supercell.
We compared the accuracy of the commonly used KEDFs. 
We found that with the WGC-KEDF, the calculated results agree quite well with the more accurate data obtained by KS-DFT calculations.
Sometimes it is not easy to obtain converged results with WGC-KEDF. 
Usually we can apply a slightly modified WGC-KEDF in which $\partial^2 w/\partial n^2 = 0$ to overcome this obstacle. 
On the other hand, the WT-KEDF tends to give a wrong estimation especially for vacancy, and is not suitable for the simulating of defect-containing systems.

Our results suggest that with carefully chosen KEDF, the accuracy of OF-DFT calculation is comparable with that of the KS-DFT calculation which is more demanding in computer resources. 
So it is promising to apply the OF-DFT method with WGC-KEDF to study large-scale systems with defects, such as the collision cascade process in irradiated materials.

\section*{Acknowledgement}

This work was supported by
the NSFC (Grant Nos. 11404299, 11305147, and 21471137),
the ITER project (No. 2014GB111006),
the National 863 Program (No. SQ2015AA0100069),
the Foundation of President of CAEP (Nos. 2014-1-58 and 2015-2-08),
and the Foundation for Development of
Science and Technology of CAEP (Grant No. 9090707).

\section*{References}
\bibliography{mybibfile}
\end{document}